\documentclass[sigconf,screen]{acmart}

\usepackage{xspace}
\usepackage{tabularx}
\usepackage{xcolor}

\newboolean{showcomments}
\setboolean{showcomments}{true} 
\ifthenelse{\boolean{showcomments}}
{\newcommand{\nb}[2]{
  \fcolorbox{black}{yellow}{\bfseries\sffamily\scriptsize#1} 
  {\sf\small$\blacktriangleright$\textit{#2}$\blacktriangleleft$}
 }
 
}
{\newcommand{\nb}[2]{} 
 
}

\newboolean{showcolor}
\setboolean{showcolor}{true}
\ifthenelse{\boolean{showcolor}}
{}
{}

\newcommand{\gh}{\textsc{GitHub}\xspace}
\newcommand{\hf}{\textsc{Hugging} \textsc{Face}\xspace}
\newcommand{\hfc}{\textsc{HFCommunity}\xspace}

\AtBeginDocument{%
  \providecommand\BibTeX{{%
    \normalfont B\kern-0.5em{\scshape i\kern-0.25em b}\kern-0.8em\TeX}}}

\setcopyright{acmlicensed}
\copyrightyear{2024}
\acmYear{2024}
\acmDOI{XXXXX.XXXXX}

\acmConference[ESEM'24]{ACM / IEEE International Symposium on Empirical Software Engineering and Measurement}{October 20--25, 204}{Barcelona, Spain}
\acmISBN{978-1-4503-XXXX-X/18/06}

\begin{document}



\title{On the Creation of Representative Samples of Software Repositories}

\author{June Gorostidi} 
\email{jgorostidie@uoc.edu}
\orcid{0009-0005-8867-0643}
\affiliation{%
  \institution{IN3 -- UOC}
  \city{Barcelona}
  \country{Spain}
}

\author{Adem Ait}
\email{adem.ait@uni.lu}
\orcid{0000-0002-5334-9041}
\affiliation{%
  \institution{University of Luxembourg}
  \city{Esch-sur-Alzette}
  \country{Luxembourg}
}

\author{Jordi Cabot}
\email{jordi.cabot@list.lu}
\orcid{0000-0003-2418-2489}
\affiliation{%
  \institution{Luxembourg Institute of Science and Technology}
  \city{Esch-sur-Alzette}
  \country{Luxembourg}
}

\author{Javier Luis Cánovas Izquierdo}
\email{jcanovasi@uoc.edu}
\orcid{0000-0002-2326-1700}
\affiliation{%
  \institution{IN3 -- UOC}
  \city{Barcelona}
  \country{Spain}
}

\renewcommand{\shortauthors}{}

\begin{abstract}
Software repositories is one of the sources of data in Empirical Software Engineering, primarily in the Mining Software Repositories field, aimed at extracting knowledge from the dynamics and practice of software projects.
With the emergence of social coding platforms such as \textsc{GitHub}, researchers have now access to millions of software repositories to use as source data for their studies. 
With this massive amount of data, sampling techniques are needed to create more manageable datasets. 
The creation of these datasets is a crucial step, and researchers have to carefully select the repositories to create representative samples according to a set of variables of interest.
However, current sampling methods are often based on random selection or rely on variables which may not be related to the research study (e.g., popularity or activity).
In this paper, we present a methodology for creating representative samples of software repositories, where such representativeness is properly aligned with both the characteristics of the population of repositories and the requirements of the empirical study.
We illustrate our approach with use cases based on \textsc{Hugging Face} repositories. 
\end{abstract}




\maketitle

\section{Introduction}
\label{sec:introduction}
Empirical software engineering is usually defined as the study of software engineering following an empirical method, which includes case studies, and different types of measurement and analysis~\cite{rombach1993experimental}.
One crucial part to enable the empirical method is the provision of the empirical data to be analyzed.

Software repositories have traditionally been one of the main sources of data in empirical software engineering, primarily in the Mining Software Repositories (MSR) field, aimed at extracting knowledge from the dynamics and practice of software projects.
With the emergence of Open-Source Software, empirical studies have mainly relied on social coding platforms, being \gh the most representative one, with more than 80 million users and 200 million repositories.
This large number of repositories makes unfeasible the analysis of the whole population and prompts sampling as a mandatory process in empirical studies.

When building samples, its representativeness, that is, how well the sample resembles the population of interest, is key to ensure the quality of the study. 
Representativeness is usually measured according to one or more variables of interest, which should be the relevant variables of the empirical study.
However, guaranteeing representativeness is hard, as researchers should consider methodological aspects and constraints such as the type and range of the variables, the composition of variables or the use of stratified methods.\looseness-1

Current approaches do not provide clear methods for creating representative samples of software repositories. 
For instance, Cosentino et al.~\cite{DBLP:conf/msr/CosentinoIC16} performed a systematic mapping study and found that few works applied probabilistic sampling.
Moreover, they revealed that stratified random sampling, which takes into account project and user diversity~\cite{DBLP:conf/sigsoft/NagappanZB13}, is only applied in just 3.2\% of the works.
Only a few studies use random sampling (e.g.,~\cite{DBLP:conf/icse/AmirR18,DBLP:journals/ese/BaltesR22}), but their samples are either too small to assume representativeness or drawn conclusions from biased sampling frames~\cite{DBLP:journals/ese/BaltesR22}. 
Ayala et al.~\cite{DBLP:journals/tse/AyalaTFJ22} noticed that most of the analyzed mining software repository studies did not select their repositories following any random sampling. 
In fact, a common approach relies on a single variable for sampling (e.g., number of stars or followers) and selects a number of the top repositories according to that variable, thus hampering the representativeness of the sample.

In this paper, we propose a four-step method to create representative samples of software repositories relying on the variables of interest and the characteristics of the repositories.
We illustrate our approach with use cases based on Hugging Face repositories.
We provide an implementation of the approach which also includes a replicability package.

The paper is structured as follows.  
Section~\ref{sec:background} presents the background. Section~\ref{sec:approach} describes our approach. 
Section~\ref{sec:related} presents the related work, and Section~\ref{sec:conclusions} concludes the paper.

\section{Background}
\label{sec:background}
In this section, we present the concepts of sampling, representativeness, and sampling of software repositories. 
We end the section with a running example.

\subsection{Sampling}
\label{sec:background:sampling}

Sampling is the process of systematically selecting a subset of elements from a large population to make inferences about the entire population.
In most sampling situations, it is crucial to assess the accuracy or confidence associated with the inferences. 
Common measures used to evaluate accuracy include confidence intervals or margin of error, each offering valuable insights into the reliability of the inferences.

Stratified sampling is a technique broadly used to improve precision and promote representativeness in the inferences~\cite{lohr2021sampling}.
In stratified sampling, the population is partitioned into independent regions or strata, and a sample is then done within each stratum.
Since strata sampling is made independently, the variances of inference results for individual strata can be added to obtain the variance for the whole population.
Furthermore, as only the within-stratum variances enter into the variances of estimators, the principle of stratification is to partition the population in such a way that the elements within a stratum are as similar as possible.
Stratified sampling is preferred when subgroups of the population may have different mean values for the targeted variables~\cite{lohr2021sampling}.

If performed correctly, stratified sampling will provide precise (i.e., low variance) estimates for the whole population~\cite{lohr2021sampling,thompson2012sampling}. 
Note that stratified sampling is usually applied to the full set of variables under study, but in the context of this work, we apply stratification to the variables of interest.

\subsection{Representativeness}
\label{sec:background:representativeness}
The concept of representativeness has been discussed in previous works~\cite{DBLP:journals/ese/BaltesR22}.
We say that a sample is representative when each sampled element represents the variables of a known number of elements in the population.

To ensure representativeness, the sampling strategy must be robust against~\cite{lohr2021sampling}: (1) selection bias, (2) measurement bias, and (3) sampling and non-sampling errors.
Selection bias occurs when some part of the target population is not in the sampled population. 
Measurement bias occurs when the measuring instrument tends to differ from the true value of the population. 
Sampling errors refer to the variability introduced by the random selection process, while non-sampling errors are any errors that cannot be attributed to the sample-to-sample variability (e.g., missing or invalid values).

To measure the sampling error, two parameters are used~\cite{lohr2021sampling,thompson2012sampling}: (1) margin of error and (2) confidence level. 
The margin of error $\epsilon$ represents the maximum acceptable difference between the sample estimate and the true population value (i.e., smaller $\epsilon$ implies greater precision or representativeness).
The confidence level, usually expressed as a percentage, indicates the degree of certainty that the true value of the population parameter lies within the $\epsilon$ (i.e., higher confidence levels ensure that the results are reliable). 

\subsection{Sampling Software Repositories}
\label{sec:background:samplingSoftware}
When performing empirical software studies in the MSR field, the source of data is usually social coding platforms (e.g., \gh or \hf).
As these platforms host thousands, or even millions, of repositories, the amount of data to perform empirical studies is massive and the use of sampling techniques is mandatory.
Furthermore, sampling should address the variables of interest in the study.
For instance, a study on repository types of \hf (i.e., datasets, models or spaces) would require building a sample with a representative number of repositories of each type.
However, current sampling methods are often based on random selection and/or rely on variables which may not be related to the study (e.g., top-most liked repositories), thus hampering the representativeness of the sample.

Sampling strategies in software repositories are inherently robust against selection and measurement bias, as we have access to the entire population and the absence of a measurement instrument, respectively.
However, sampling errors and non-sampling errors can still occur and must be handled appropriately.
The former can be measured using $\epsilon$ and confidence levels for each variable, while the latter mainly appears as Not a Number (NaN) values. 
By carefully selecting $\epsilon$ and dealing with NaN values, we can ensure that our sample represents the whole population within a range of certainty.\looseness-1

\subsection{Running Example}
To illustrate our approach, we will use a running example based on \hfc~\cite{DBLP:journals/scp/AitIC24}, an offline database built from the data available at the \textsc{Hugging Face Hub} (HFH).
HFH is a social coding platform to host machine leargning models, datasets and spaces (i.e., repositories aimed at showcasing models).
These artifacts are hosted as Git repositories, with additional social features such as discussion threads.
Since its creation, HFH has been rapidly growing up to the number of more than 950k public repositories by April 2024. 

\begin{figure}[t]
    \centering
    \includegraphics[width=\columnwidth]{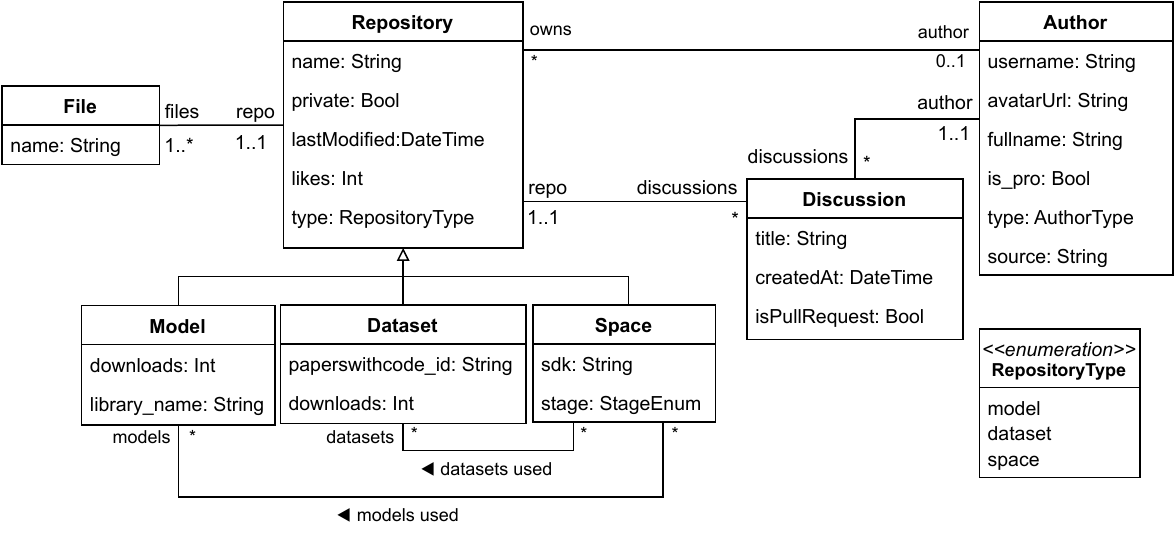}
    \caption{Part of the conceptual schema of \hfc.}
    \label{fig:reducedSchema}
\end{figure}

\hfc collects data from HFH and Git repositories, and stores it in a relational database to facilitate their analysis.
\hfc provides both HFH and Git history data as a SQL dump, periodically released.
Figure~\ref{fig:reducedSchema} shows a snippet of the conceptual schema of \hfc.

As running example, we will use three use cases based on \hfc repositories, to build representative samples based on (1) one numerical variable, (2) one categorical variable, and (3) two variables of different type.

\section{Our approach}
\label{sec:approach}
\begin{figure}[t]
    \centering
    \includegraphics{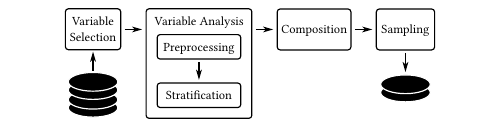}
    \caption{Our approach.}
    \label{fig:approach}
\end{figure}

We have developed a methodology employing stratified random sampling to create representative samples of software repositories, focusing on key variables of interest. 
Figure~\ref{fig:approach} illustrates our approach.
The process is split into four phases: (1) variable selection, (2) variable analysis, (3) composition, and (4) sample creator.
Variable analysis is further divided into two steps, namely: preprocessing and stratification.
Next, we describe each step and illustrate their application in the three use cases of our running example.
Finally, we describe the implemented tool support.

\subsection{Variable Selection}
This phase identifies the variables which will drive the creation of the sample.
The selection process must carefully be performed, as the resulting sample will be representative of the population based on these key variables.
Note that the resulting sample will include both the selected variables and the remaining variables of the population dataset.

\smallskip
\noindent\textbf{Case \#1}.
We will build a sample using the numerical variable \texttt{likes} of the \texttt{Repository} class, which indicates the number of likes for the repository.

\smallskip
\noindent\textbf{Case \#2}.
We will build a sample using the categorical variable \texttt{type} of the \texttt{Repository} with three possible values (i.e., dataset, model, and space).

\smallskip
\noindent\textbf{Case \#3}.
We will build a sample using the variables \texttt{likes} and \texttt{type}, thus mixing variables of different types.

\subsection{Variable Analysis}
This phase performs the analysis of the selected variables.
We consider two types of variables: numerical and categorical.
The numerical variables refer to data that is measured on a continuous scale (e.g., number of likes or downloads). 
On the other hand, categorical variables refer to data that can be divided into groups (e.g., programming languages used in the code or the license).
The analysis process includes two steps: preprocessing and stratification.
Next, we describe each step and illustrate the application in the three use cases.

\subsubsection{Preprocessing}
This step focuses on studying the descriptive characteristics (e.g., range, distribution, etc.) of the variable.
In particular, a key aspect is to handle non-sampling errors (i.e., missing or NaN values), as commented in Section~\ref{sec:background:samplingSoftware}. 
For numerical variables, they can be imputed or removed.
Imputation is the process of replacing missing data with calculated values (e.g., mean or median).
While the imputation is desired in those cases where all the observations are required for the analysis, removal is applied when these values do not provide any information.
When sampling software repositories, in most cases the suitable solution is to remove those observations where the numerical variable has a NaN value, as imputing them usually introduces false information.
Once the NaN/missing values have been addressed, we will proceed to calculate essential statistical metrics such as the population size, mean, and standard deviation.
These calculations will allow us to compare these metrics with the corresponding values derived from our sample.

For categorical variables, we do not exclude NaN values; instead, we classify NaN values as an additional category and include them in the sampling process. 
NaN values in categorical variables can carry meaningful information about missing data patterns, and their inclusion ensures that the sample accurately represents all aspects of the population.
For instance, given a categorical variable indicating the programming languages used in a repository, a missing or NaN value may mean either the repository does not include code or that no language has been reported.
Furthermore, since categorical variables do not require the calculation of statistical values (e.g., for imputing), treating NaN values as another category does not affect representativeness.

\smallskip
\noindent\textbf{Case \#1}.
The variable \texttt{likes} comprises 674,827 observations, which corresponds to the whole set of repositories in \hfc.
The distribution of this variable is very skewed, as the majority of values are either 0 or close to 0.
As there is little consensus when reporting descriptive statistics for skewed data, we report median, Inter-quartile Range (IQR), average, and standard deviation.
A repository has a median number of 0.0 likes and IQR of (0,0) ($\mu = 1.13$, $\sigma = 28.13$). 

\begin{figure}[t]
    \centering
    \includegraphics[width=0.9\columnwidth]{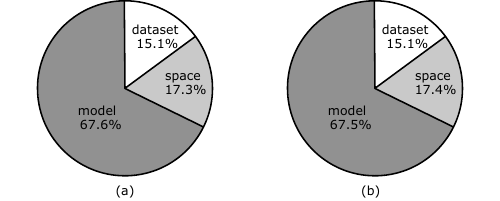}
    \caption{Data Distribution of the sample for \texttt{type} variable. (a) Original and (b) sample data distribution.}
    \label{fig:case2:distribution}
\end{figure}

\smallskip
\noindent\textbf{Case \#2}.
The variable \texttt{type} has a total of 674,827 observations, from which there are 456,303 models, 101,681 datasets, and 116,843 spaces, with no missing values. 
Figure~\ref{fig:case2:distribution}a shows the distribution of the categories. 

\smallskip
\noindent\textbf{Case \#3}. 
The variables considered in this use case have been already described in the previous cases.

\subsubsection{Stratification}
\label{sec:approach:stratification}
The stratification process depends on the variable type.
For numerical variables, we propose to apply a clustering algorithm to define the strata.
We suggest the K-means algorithm, which is widely used in data mining, but other clustering algorithms may be considered according to the data characteristics.
The K-means algorithm clusters data into $K$ groups based on data similarities, iteratively assigning observations to clusters that minimize variance within each cluster and maximize differences between clusters.

We have a population divided into K disjoint strata.
Given a finite population of size $N$, the sizes of the different strata, $N_1, N_2, \ldots, N_K$, are known, where $N = \sum_{i=1}^{K} N_i$.
The proportion of elements of the population within the $i$-th stratum is $\phi_i = \frac{N_i}{N}$. In stratum $i$, the mean of the variable of interest is $\mu_i$ and its standard deviation is $\sigma_i$.
The population mean $\mu$ can be expressed as listed in Equation~\ref{eq:populationMean}, which is the weighted mean of the means within each stratum.

\begin{equation}\label{eq:populationMean}
    \mu = \sum_{k=1}^{K} \phi_k\mu_k
\end{equation}

For each stratum $k$, we build a sample of size $n_k$, where the estimator of the mean in such stratum is shown in Equation~\ref{eq:meanEstimator2}.

\begin{equation}\label{eq:meanEstimator2}
    \overline{x}_k = \frac{1}{n_k} \sum_{i=1}^{n_k} x_{k_{i}}
\end{equation}

Substituting the mean $\mu_k$ of each stratum by its estimator $\overline{x}_k$ in Equation \ref{eq:populationMean}, we get the estimated population mean, or sample mean (see Equation~\ref{eq:meanEstimatorStrata}).

\begin{equation}\label{eq:meanEstimatorStrata}
    \overline{x} = \sum_{k=1}^{K} \phi_k\overline{x}_k
\end{equation}

Assuming that the observations within each stratum are made independently, the variance of the estimated population mean is defined in Equation~\ref{eq:variance}.

\begin{equation}\label{eq:variance}
    Var(\overline{x}) = \sum_{k=1}^{K} \phi_k^2 Var(\overline{x}_k)
\end{equation}

Since our population is finite, $\text{Var}(\overline{x}_k) = \frac{\sigma^2_k}{n_k} \left(1 - \frac{n_k}{N_k}\right)$, where $\sigma^2_k$ is the variance of the variable of interest in the $k$-th stratum.
Equation~\ref{eq:varPobFinita} shows the variance of the estimated population mean in this case.

\begin{equation}\label{eq:varPobFinita}
    Var(\overline{x}) = \sum_{k=1}^{K} \phi_k^2 \frac{\sigma^2_k}{n_k} \left(1 - \frac{n_k}{N_k}\right)
\end{equation}

If the variance $\sigma^2_k$ is not known, it can be estimated using the corresponding sample variance for the $k$-th stratum $s^2_k$.
Using $s^2_k$, the variance of the estimated population mean, or variance of the sample mean, would be obtained from the previous Equation~\ref{eq:varPobFinita}, expressed in Equation~\ref{eq:se2}. 


\begin{equation}\label{eq:se2}
    s^2_{\overline{x}} = \sum_{k=1}^{K} \phi_k^2 \frac{s^2_k}{n_k} \left(1 - \frac{n_k}{N_k}\right)
\end{equation}

Equation~\ref{eq:meanEstimator} shows the confidence interval for the population mean with large $n$, where $z_{1-\alpha/2}$ is the critical value of the standard normal distribution for a specific level of confidence, and $s_{\overline{x}}$ is the standard deviation of the sample mean.

\begin{equation}\label{eq:meanEstimator}
    \mu \in \left[ \overline{x} \pm z_{1-\alpha/2}s_{\overline{x}} \right]
\end{equation}

To estimate $\mu$ with an error no greater than $\epsilon$ and a confidence of $1-\alpha$, the sample size is obtained by solving the Equation~\ref{eq:epsilon}.

\begin{equation}\label{eq:epsilon}
    \epsilon = z_{1-\alpha/2}s_{\overline{x}}
\end{equation}

To calculate the sample size $n$, we rely on the Equation~\ref{eq:epsilon}. 
As our population is of finite size $N$, considering Equation~\ref{eq:se2}, we can obtain Equation~\ref{eq:f1}.

\begin{equation}\label{eq:f1}
\epsilon = z_{1-\alpha/2} \sqrt{\sum_{i=1}^{K} \phi_2^2 \frac{s_i^2}{n_i} \left( 1 - \frac{n_i}{N_i}\right)}
\end{equation}

To solve this equation, we first need to decide how to distribute the sample size $n$ into the sample sizes $n_1, n_2, \ldots, n_K$ for each stratum, using an allocation method.
Being $n_i = n \cdot w_i$ the allocation method used to distribute the sample, we have the results shown in Equation~\ref{eq:f2}.

\begin{align}\label{eq:f2}
\epsilon &= z_{1-\alpha/2}\sqrt{\sum_{i=1}^{K} \phi_i^2 \frac{s_i^2}{n w_i} \left( 1-\frac{n w_i}{N_i}\right)} \nonumber \\
 &= z_{1-\alpha/2} \sqrt{\sum_{i=1}^{K} \phi_i^2 s_i^2 \left( \frac{1}{n w_i} - \frac{1}{N_i} \right)}
\end{align}

If we isolate $n$ we obtain the result shown in Equation~\ref{eq:f3}.

\begin{equation}\label{eq:f3}
n = \frac{\sum_{i=1}^{K} \frac{\phi_i^2 s_i^2}{w_i}}{\left( \frac{\epsilon}{z_{1-\alpha/2}} \right)^2 + \sum_{i=1}^{K} \frac{\phi_i^2 s_i^2}{N_i}}
\end{equation}

There are three commonly used allocation methods: (1) uniform allocation, (2) proportional allocation, and (3) optimal allocation.
We use the proportional allocation method, as it distributes the sample size proportionally across the different strata within the population.
This method ensures that each stratum is adequately represented in the final sample, thus enabling the results to be generalized to the entire population~\cite{lohr2021sampling}.
Using the proportional allocation method, Equation~\ref{eq:proportional} calculates the size of the sample for stratum $i$.
Note that the value of $n_i$ is rounded up.

\begin{equation}\label{eq:proportional}
n_i = n \cdot \frac{N_i}{N} = n \cdot \phi_i
\end{equation}

Using the proportional allocation method, $w_i = \phi_i$ in $n_i = n \cdot w_i$ before presented. Thus, Equation~\ref{eq:f3} becomes the Equation~\ref{eq:f3phi}.

\begin{equation}\label{eq:f3phi}
n = \frac{\sum_{i=1}^{K} \frac{\phi_i^2 s_i^2}{\phi_i}}{\left( \frac{\epsilon}{z_{1-\alpha/2}} \right)^2 + \sum_{i=1}^{K} \frac{\phi_i^2 s_i^2}{N_i}}
\end{equation}

Considering that in finite populations $\phi_i = \frac{N_i}{N}$, the previous Equation~\ref{eq:f3phi} reduces to Equation~\ref{eq:f4}.

\begin{equation}\label{eq:f4}
n = \frac{\sum_{i=1}^{K} \frac{N_i}{N} s_i^2}{\left( \frac{\epsilon}{z_{1-\alpha/2}} \right)^2 + \frac{1}{N} \sum_{i=1}^{K} \frac{N_i}{N} s_i^2}
\end{equation}

To obtain $n$, we have to define the margin of error $\varepsilon$ and set the confidence level $1-\alpha$. 
The choice of $\epsilon$ must be customized for each variable since each one has a unique range, and there is no universal value applicable to all. 
We propose to set $\epsilon$ using the mean of each variable. 
Once the $\epsilon$ is set, a specific $\epsilon$ for each variable is calculated by multiplying the mean by this value, that is, $\epsilon = \mu_{\text{variable}} \cdot \epsilon_{\text{desired}}$.

For categorical variables, this step is straightforward, as the strata are inherently defined by the categories of the variable and each category has a proportion $\hat{p}$ within the population (see Equation~\ref{eq:c1}).

\begin{equation}\label{eq:c1}
    \hat{p} \pm z_{1-\alpha/2} \sqrt{\frac{\hat{p}(1-\hat{p})}{n}}
\end{equation}

We can directly select the desired sample size and perform the sampling respecting the category proportions ($\hat{p}$) or calculate the sample size given a desired $\epsilon$ and a confidence interval.
The $\epsilon$ for categorical variables is then calculated as shown in Equation~\ref{eq:c2}.

\begin{equation}\label{eq:c2}
    \epsilon = z_{1-\alpha/2} \sqrt{\frac{\hat{p}(1-\hat{p})}{n}}
\end{equation}

As before, we can isolate \textit{n} from Equation~\ref{eq:c2} to calculate the size of the sample and obtain Equation~\ref{eq:c3}.

\begin{equation}\label{eq:c3}
    n = \frac{\hat{p}(1-\hat{p})}{\left( \frac{\epsilon}{z_{1-\alpha/2}} \right)^2}
\end{equation}

Since we have multiple categories in each variable, to avoid calculating different sample sizes for each category value, it is possible to apply $\hat{p} = 0.5$, which maximizes the value of $\epsilon$.
Thus, substituting the value of $\hat{p} = 0.5$ in Equation~\ref{eq:c2}, we obtain a generalized formula in Equation~\ref{eq:c4}.

\begin{equation}\label{eq:c4}
    n = \left( \frac{z_{1-\alpha/2}}{2\epsilon} \right)^2 
\end{equation}

\smallskip
\noindent\textbf{Case \#1}.
To perform the stratification, we apply the k-means clustering algorithm to generate the strata, leveraging the elbow method to determine the optimal number of clusters, which is 3.
Next, we need to calculate the number of elements in each stratum applying Equation~\ref{eq:f4}.
We also calculate the proportions of each segment relative to the entire population, and the standard deviations for each stratum.
We apply the proportional allocation approach, use an $\epsilon$ of 0.114 and a confidence interval of 0.95.
By employing these parameters along with the function governing sample size determination, we derive that for the \texttt{likes} variable, the overall sample size must be 57,258 observations, and each of the 3 stratum will have a size of 57,239, 18 and 1 observations.

\smallskip
\noindent\textbf{Case \#2}.
Since categorical variables define strata through their categories, we can directly visualize the population distribution in Figure~\ref{fig:case2:distribution}a without the need for any additional clustering algorithms.
In this scenario, we need to calculate the proportions of the strata. 
With an $\epsilon$ of 0.05 and a confidence interval of 0.95, a sample size of 385 was determined. 

\smallskip
\noindent\textbf{Case \#3}.
The identification of the strata for \texttt{likes} and \texttt{type} variables have already been described in the previous cases.

\subsection{Composition}
\label{sec:approach:composition}
This phase involves composing the strata of the selected variables to form a new strata distribution.
To this aim, we first generate all possible combinations by means of a permutation process, and then we select the valid ones.
A combination of strata is valid when it includes at least one member with observations for each variable in the combination.
Although several variables can be used for the composition, we recommend using no more than four to six variables, as the probability of some variables canceling the effects of others increases with the number of variables.

\smallskip
\noindent\textbf{Cases \#1 \& \#2}.
As these use cases only consider one variable, there is no need for the composition phase.

\smallskip
\noindent\textbf{Case \#3}.
We are composing a numerical variable with a categorical one, so we generate all possible combinations.
Each variable has three strata.
For the \texttt{likes} variable, the strata are defined by the following ranges: $[0, 315]$, $[2285, 9909]$, and $[317, 1930]$.
For the \texttt{type} variable, the strata include repositories of type \texttt{dataset}, \texttt{model}, and \texttt{space}. 
By generating all possible combinations, we obtain 9 strata as reported in the four first columns of Table~\ref{tab:case3:composition}.

\begin{table}[t]
    \caption{Composition and sampling process for \texttt{likes} and \texttt{type} variables.}
    \label{tab:case3:composition}
    \footnotesize
    \begin{tabular}{ccrrcr}
        \multicolumn{4}{c}{\textsc{Composition}} & &
        \multicolumn{1}{c}{\textsc{Sampling}} \\
    \cmidrule{1-4} \cmidrule{6-6}
        \multicolumn{1}{c}{Stratum} & 
        \multicolumn{1}{c}{\texttt{type}} & 
        \multicolumn{1}{c}{\texttt{likes}} & 
        \multicolumn{1}{c}{n} & &
        \multicolumn{1}{c}{n} \\
    \cmidrule{1-4} \cmidrule{6-6}
        1 & \texttt{dataset} & $[0, 437]$     & 101,667 & &  58 \\ 
        2 & \texttt{dataset} & $[442, 2597]$  & 13       & &   0 \\ 
        3 & \texttt{dataset} & $[2691, 9909]$ & 1      & &   0 \\ 
        4 & \texttt{model}   & $[0, 437]$     & 456,149 & & 260 \\ 
        5 & \texttt{model}   & $[442, 2597]$  & 143      & &   0 \\ 
        6 & \texttt{model}   & $[2691, 9909]$ & 11     & &   0 \\ 
        7 & \texttt{space}   & $[0, 437]$     & 116,804 & &  67 \\ 
        8 & \texttt{space}   & $[442, 2597]$  & 38       & &   0 \\ 
        9 & \texttt{space}   & $[2691, 9909]$ & 1      & &   0 \\ 
    \bottomrule
    \end{tabular}
\end{table}

\subsection{Sampling}
\label{sec:approach:sampling}
For numerical variables, we apply a probabilistic sampling method such as simple random sampling to collect the observations for each stratum. 
We also calculate the statistical parameters to enable a comparison with those of the population. 
To ensure the quality of the sample, the process can be applied iteratively, typically five to ten times, to select the best sample according to the statistical parameters.

For categorical variables, we extract samples from each stratum, applying simple random sampling and maintaining their proportions relative to the population. 
As commented in Section~\ref{sec:approach:stratification}, we can calculate the sample size either using the formula with a given $\epsilon$ and confidence interval, or directly choosing a sample size but respecting the category proportions.
In the former case, note that there may be a slight discrepancy in the sample proportions.

When mixing both numerical and categorical variables, we apply the same process applied for categorical variables, but for each stratum of the resulting combination.

\smallskip
\noindent\textbf{Case \#1}.
To perform the sampling of the \texttt{likes} variable, we apply iteratively the simple random sampling method to extract samples from each stratum.
We compute the estimation of the mean, the standard error of the mean, and the confidence intervals to assess the sample quality.
Table~\ref{tab:case1:sampling} shows the results of the sampling process using 5 iterations, being the iteration 3 the most appropriate.

\begin{table}[t] 
    \caption{Sampling process for the \texttt{likes} variable.}
    \label{tab:case1:sampling}
    \footnotesize
    \begin{tabular}{crrrrr}
        \multicolumn{1}{c}{it.} & 
        \multicolumn{1}{c}{N} & 
        \multicolumn{1}{c}{n} & 
        \multicolumn{1}{c}{$\mu$} & 
        \multicolumn{1}{c}{$\overline{x}$} & 
        \multicolumn{1}{c}{C.I.} \\ 
    \toprule
        1 & 674,827 & 55,498 & 1.13 & 1.10 & [0.99, 1.21] \\
        2 & 674,827 & 55,498 & 1.13 & 1.15 & [1.05, 1.27] \\
        3 & 674,827 & 55,498 & 1.13 & 1.13 & [1.02, 1.24] \\
        4 & 674,827 & 55,498 & 1.13 & 1.16 & [1.01, 1.27] \\
        5 & 674,827 & 55,498 & 1.13 & 1.11 & [1.00, 1.22] \\
    \bottomrule
    \end{tabular}
\end{table}

\smallskip
\noindent\textbf{Case \#2}.
For the \texttt{type} variable, we conducted the sampling process for each stratum.
The sample has 385 observations, from which there are 260 models, 58 datasets, and 67 spaces.
Figure~\ref{fig:case2:distribution}b shows the distribution of the categories.
Note that the distribution in the sample is similar to the population distribution (cf. Figure~\ref{fig:case2:distribution}a).

\smallskip
\noindent\textbf{Case \#3}.
The sampling process for the \texttt{likes} and \texttt{type} variables builds 3 strata of a total size of 385 observations.
The last column of Table~\ref{tab:case3:composition} shows the size of each stratum.

\subsection{Tool Support}
\label{sec:approach:toolSupport}
We have implemented in Python the corresponding tool support\footnote{\url{https://github.com/SOM-Research/sample-creator}}, which includes the four phases of the methodology and liberates the user from dealing with the mathematical foundations.
The tool repository also includes a reproducibility and replicability package for the use cases presented.

\section{Related Work}
\label{sec:related}
Sampling in empirical software engineering is a crucial step to build datasets.
Nagappan et al.~\cite{DBLP:conf/sigsoft/NagappanZB13} noticed the need for sampling due to the increasing number of Open-Source projects and proposed a technique to assess how well a study covers a population of projects.
Dabic et al.~\cite{DBLP:conf/msr/DabicAB21} presented a dataset to facilitate the sampling of \gh repositories, based on frequently used project selection criteria in MSR studies.
In this paper, we aim at facilitating the creation of these samples.

Despite their relevance, the application of sampling methods is far from normalized, as noted by the meta-analysis by Cosentino et al.~\cite{DBLP:conf/msr/CosentinoIC16}.
Other works have also noticed that the use of random sampling is scarce~\cite{DBLP:conf/icse/AmirR18,DBLP:journals/ese/BaltesR22,DBLP:journals/tse/AyalaTFJ22}.

Given this situation, there are works aimed at studying the task of sampling in software engineering.
Baltes et al.~\cite{DBLP:journals/ese/BaltesR22}, conducted a critical review to investigate how sampling methods are used in software engineering.
Besides addressing the common misunderstanding of the representativeness term, they provide a set of guidelines for conducting, reporting and reviewing sampling.
However, they do not provide a methodology, as we present in this paper.
Castaño et al.~\cite{DBLP:journals/corr/abs-2402-07323} delved into the tools and methodologies used in their previous studies of HFH, and propose dimensions to consider when performing sampling of HFH models, which might help in selecting variables for sampling, but do not provide any method to follow.
Further studies~\cite{DBLP:conf/icse/TorchianoFTM17,DBLP:conf/cibse/MelloT15,DBLP:conf/esem/MelloT16} provide recommendations for sampling in non-MSR studies. 
However, to the best of our knowledge, there is no work providing a method to create representative samples given a set of variables of interest in software repositories.


\section{Conclusions}
\label{sec:conclusions}
In this paper, we have presented a methodology for creating representative samples of software repositories.
Our approach allows selecting one or more variables of interest related to software repositories to build a sample that is representative for such variables.
The approach provides guidance for both numerical and categorical variables, and its composition, if required.
We have illustrated our approach with three use cases based on \hf repositories using the \hfc dataset, and provided a tool implementation of the approach, including a replicability package.

As future work, we plan to extend our approach with support for sampling from multiple datasets or for evolving datasets (i.e., adding or removing observations or variables).
Current implementation is focused on building samples given an $\epsilon$ and a confidence level, which ensures high representativeness but may result in large samples.
We plan to allow the sample size parameterization and report how representativeness may be affected. 
Finally, we are also interested in applying our approach to additional use cases. 

\begin{acks}
This work is part of the project TED2021-130331B-I00 funded by MCIN/AEI/10.13039/501100011033 and European Union NextGenerationEU/PRTR; and BESSER, funded by the Luxembourg National Research Fund (FNR) PEARL program, grant agreement 16544475.
\end{acks}

\bibliographystyle{ACM-Reference-Format}
\bibliography{esem24-bib}

\end{document}